\input preprint.sty

%
\title{Algebraic Techniques for Enumerating Self-Avoiding
Walks on the Square      Lattice}

\author{ A R Conway\dag, I G Enting\dag\ddag\  and A J Guttmann\S\P}

\address{\dag\ Department of Mathematics, The  University of Melbourne,
Parkville, Vic.~3052, Australia}

\address{\ddag\ Permanent address: CSIRO, Division of Atmospheric Research,
Private Bag 1, Mordialloc, Vic.~3195, Australia}

\address{\S\ Department of Theoretical Physics, Oxford University,
1 Keble Road, Oxford, OX1 3NP,U.K.}

\address{\P\ Permanent Address:  Department of Mathematics, The
University of Melbourne, Parkville, Vic.~3052, Australia}

\shorttitle{Self-avoiding walks}


\jnl{Journal of Physics A}

\date

\beginabstract
We describe a new algebraic technique for enumerating self-avoiding walks
on the
rectangular lattice.  The computational complexity of enumerating walks of
$N$ steps is
of order $3^{N/4}$ times a polynomial in $N$, and so the approach is greatly
superior to
direct counting techniques.  We have enumerated walks of up to 39 steps.
As a consequence, we are able to accurately estimate the critical point,
critical exponent, and critical amplitude.

\endabstract

\section{Introduction}

Over the years, the enumeration of square lattice self-avoiding walks has
become a benchmark, first for computer performance, and more recently for
algorithm design. In the early `70s, Sykes et al. (1972) obtained 24 terms of
the series by using the chain counting theorem. Direct enumeration is
probably somewhat faster, but the graphs enumerated for the chain
counting theorem (figure-eights, theta graphs, dumbells and
polygons) were useful for other problems in the theory of phase transitions,
most notably the Ising model. In 1987 Guttmann extended the series by
three terms, using direct enumeration, and in 1991 Guttmann and Wang
using a dimerisation algorithm obtained two further terms. Subsequently
McDonald et al. (1992) obtained a further term, using an extension of
dimerisation to trimerisation. Late in 1991, Masand et al. (1992)
used a large supercomputer, a CM-2 containing 65536 processors, to extend the
series to 34 terms, running for $\sim 100$ hours. All these advances came
about because
of improvements in computer technology (in large part), with relatively small
improvements brought about by algorithm design. (Dimerisation or trimerisation
saves a factor of around 2 or 3, but does not ameliorate the exponential
growth rate of computer time).

The finite-lattice method plus transfer matrices described here allows
35 terms to be obtained on a work station (an IBM 6000/530 with 256MB of
memory) in less time than the 65536 processor CM-2 took to obtain 34 terms.
Because of memory requirements, it was necessary to move to a larger
machine (an IBM 3090/400 with 500MB memory and 2GB of backing storage) to
go from 35 to 39 steps.
This computer would be capable of extending the series to 43 terms.
However that calculation might take up to a month of c.p.u. time, and
so has not been pursued. This improvement however is due to
the exponential improvement in algorithm design, rather than evolution
of computer speed. This is discussed in
more    detail below.

The method used is based on the method which Enting and Guttmann
have used extensively over the past twelve years to enumerate
 self-avoiding rings on a square lattice, but is significantly more
complicated due to requirement for a second stage of processing.

In a series of papers we have reported some significant improvements in the
enumeration of
self-avoiding rings on the square lattice extending the known series (Sykes
et al., 1972)
from 26 steps to 38 steps (Enting, 1980), 46 steps (Enting and Guttmann,
1985) and 56
steps (Guttmann and Enting, 1988).  The extension from 38 steps to 56 steps
reflects our
use of increasingly powerful computing systems and in particular the use of
increasingly
large amounts of physical and virtual memory.  Only minor changes to our
programs have
been made, primarily to `tune' the procedure to make efficient use of
particular computer
architectures.  We have also generalised our method to enumerate caliper
moments of
self-avoiding rings on the square lattice (Guttmann and Enting, 1988) and
also to
enumerate self-avoiding rings on the $L$ and Manhattan lattices (Enting and
Guttmann, 1985)
and the honeycomb lattice (Enting and Guttmann, 1989).
Most recently we have extended the enumeration of triangular lattice
polygons to 25 steps (as reported by Enting and Guttmann, 1990)
and then to 35 steps (Enting and Guttmann, 1992).

 While our techniques have been highly efficient for enumerating
self-avoiding
rings they are less suitable for enumerating self-avoiding walks.  The
difficulty is that
walks can span a larger lattice than rings because they are not
forced to
return.  A walk of $L$ steps can span a distance $L$ while a ring of $L$
steps can only  span a
distance of up to $L/2$.  For walks constrained by a surface our polygon
enumeration
 techniques could be generalised
 in analogy with our calculation of surface susceptibilities for
the square
lattice (Enting and Guttmann, 1980).

 The present paper presents an algebraic technique for enumerating
self-avoiding
walks on a rectangular lattice.  The basic quantity that we consider is
$C_{mn}$, the
number of walks from a given origin with $n$ steps in the $\pm x$
directions and $m$ steps
in the $\pm y$ directions.  We consider segments of walks that double back
in the $y$
direction and which can therefore be counted efficiently by transfer matrix
techniques.
The general enumeration can be expressed as a combination of such
irreducible
contributions.  We further improve the efficiency of the procedure by
restricting the
range of the index $n$ to $n \leq k$ and reconstruct $C_{mn}$ for $m + n \leq
2k + 1$ by
using the symmetry relation
$$C_{mn} = C_{nm}. \eqno(1.1)$$
 The layout of the remainder of this paper is as follows.  Section
2 describes the
way in which the generating function for self-avoiding walks can be
constructed from
irreducible contributions.  Section 3 shows the way in which these
irreducible
contributions can be constructed from generating functions for walks on
strips that can
be determined by algebraic techniques.  Section 4 describes the algorithms
for
determining the requisite generating functions and analyses the
computational complexity
of the procedure.  Section 5 describes our analysis of the singularity
structure of the
generating function for self-avoiding walks based on the 39 terms that we
have obtained.

\section{Generating functions for self-avoiding walks}

The generating function for self-avoiding walks on the rectangular lattice
is
$$C(u,w) = \sum_{m,n=0}^{\infty} C_{mn}u^{m}w^{n} \eqno (2.1)$$

 The enumeration of the coefficients $C_{mn}$ is restricted to
finite order in $m$ and/or
$n$ and we generally truncate the double series at $m+n \leq J$.  The
obvious summation
then gives us the number of walks of up to $J$ steps on the square lattice.

 Our enumeration procedure is based on considering projections of
walks onto the
$y$ axis to produce the type of diagrams shown in Figure 1.  We refer to
segments of
walks as irreducible if the projection of that segment onto the $y$ axis
has two or more
$y$ bonds in each position.  We will also classify irreducible segments by
the number of
$y$ bonds that they span.  At this point we need to refine the terminology
and
distinguish between walks which are directed graphs and chains which are
not directed.
Our aim is to enumerate walks while the transfer matrix techniques
generally enumerate
chains.  The decomposition of walks into irreducible components makes use
of both chains
and walks.
\indent Figure 2 shows the 5 distinct types of irreducible component that
we need to
consider.

 $P(u,w)$ is the generating function for walks that have no $y$
bonds.  Thus

$$P(u,w) = 1 + 2u + 2u^{2} + 2u^{3} ... = (1+u)/(1-u) \eqno(2.2)$$

 $Q(u,w)$ is the generating function for chains that are
irreducible and for which neither
end-point lies at an extremal $y$ coordinate.  We also consider subdividing
such cases
according to $m$, the number of $y$ bonds spanned by the projection and
define
$Q_{m}(u,w)$ accordingly. The two pertinent results are

$$Q(u,w) = \sum_{m=2}^{\infty} Q_{m} (u,w) \eqno (2.3)$$
and

$$Q_{m}(u,w) = O(w^{2m}) \eqno (2.4)$$
 Note that from the definition

$$Q_{0} \equiv Q_{1} \equiv 0 \eqno (2.5)$$

 $R(u,w)$ is the generating function for
 irreducible  chains with both endpoints
having the maximal $y$
coordinate.  Again we subdivide these chains according to the number of $y$
bonds spanned
and put

$$R(u,w) =  \sum_{m=1}^{\infty} R_{m} (u,w) \eqno (2.6)$$

$$R_{m}(u,w) = O(w^{2m}) \eqno (2.7)$$

 and arbitrarily define
$$R_{0} \equiv 0 \eqno (2.8)$$

 $S(u,w)$ is the generating function for irreducible chains that have
precisely one end having
the maximal $y$ coordinate.  Again

$$S(u,w) = \sum_{m=2}^{\infty} S_{m}(u,w) \eqno (2.9)$$

$$S_{m}(u,w) = O(w^{2m+1}) \eqno (2.10)$$
and from the definition

$$S_{0} \equiv S_{1} \equiv 0 \eqno (2.11)$$

 Finally $T(u,w)$ is the generating function for
 irreducible chains in which the two ends have
maximal and minimal $y$ coordinates.  We arbitrarily define

$$T_{0} \equiv 0 \eqno (2.12)$$

 so that
$$T(u,w) = \sum_{m=1}^{\infty} T_{m}(u,w) \eqno (2.13)$$

$$T_{m}(u,w) = O(w^{3m}) \eqno (2.14)$$

 When constructing the generating functions for walks we combine
the
irreducible components represented by $P,Q,R,S$ and $T$ by linking them
with single $y$
bonds.  These each contribute a factor of $w$ to the generating function
$C(u,w)$.  Two
single $y$ bonds that are adjacent in the projection are actually connected
by a type
$P$ irreducible component.  A chain whose projection is $k$ consecutive
single $y$ bonds
will have a factor of $P$ at each internal point.  Note that the
{\it walk}
generating function $P$ is required because each distinct direction along
the $x$ axis
will generate a distinct chain when combined with $y$ bonds.  If we sum
these
contributions we can consider linking components of types $R,S$ and $T$
with chains of one or more
$y$ bonds connecting type $P$ walks.  The generating function for such
chains is

$$
U(u,w)  =  w + wPw + wPwPw ... \
$$
$$
  =  w/(1 - wP(u,w))
\eqno (2.15)$$

 In the same way we can regard the overall chain as consisting of
end segments $P$, $R$ or
$S$ connected by combinations of ``reducible" parts with generating
function $U$ and
irreducible parts with generating function $T$.  The generating function
for chains
connecting irreducible end segments is thus

$$
 V(u,w)  =  U + UTU + UTUTU  + \dots
$$
$$
  =  U/(1-TU)
$$
$$
 =  w/(1-w(T+P))  \eqno (2.16)
$$

 It is now possible to express the self-avoiding walk generating
function as

$$
 C(u,w)
 =  P(u,w) + 2[Q(u,w) + 2R(u,w) + 2S(u,w) + T(u,w)]
 $$
 $$
  + 2V(u,w)[P(u,w) + 2R(u,w) + S(u,w) + T(u,w)]^{2}
 \eqno (2.17) $$

 The structure of expression (2.17) shows that to obtain an
expansion in powers of $w$, it is necessary
to obtain $Q,R,S,$ and $T$ to the requisite order.  If the
irreducible generating functions $Q_{m}, R_{m}, S_{m}$ and $T_{m}$ are
known for $m \leq M$
then $C(u,w)$ will be correct to $w^{2M+1}$ (the first incorrect term
arising from the
absences of $Q_{M+1}$ and $R_{M+1}$).  The use of the symmetry relations
(1.1) will give
$C_{mn}$ for $m+n \leq 4M+3.$
\bigskip
\section{Combining generating functions for chains on strips}
The transfer matrix techniques described in the next section produce
generating functions
for sets of walks confined to strips whose $y$ coordinates are bounded.
Subject to these
constraints, all chains are counted, not merely irreducible components.
Thus there is a
need to relate unrestricted generating functions to the restricted
generating functions
for irreducible components.  In our previous enumerations of self-avoiding
rings only
linear combinations of different classes of graph were involved and so the
restricted
generating functions were linear combinations of unrestricted generating
functions.  The
present formalism is more complicated because non-linear relations are
involved.  We
begin by considering $T_{M}^{*}(u,w)$, the generating function for chains
whose $y$
coordinates span $M$ bonds and which have one end at each $y$ extremum, that
is, $T_{M}^{*}$ is the generating function for bridges.
The sum over $M$ is denoted $T^{*}(u,w)$.  By
considering the appropriate subset of terms from (2.17) we have

$$
T^{*}(u,w) = T(u,w) + (P(u,w) + T(u,w))^{2}V(u,w).
\eqno(3.1) $$

 While this equation is formally correct, it is unsuitable for
relating $T^{*}$ to
$T$ because of the fact that while $T_{m}$ is of order $w^{3m}$, $T^{*}_m$ is
of order
$w^{m}$.  Thus to obtain $T$ correct to $w^{K}$ would require the
calculation of
$T^{*}_{m}$ for $m \leq K.$  This difficulty is avoided by introducing an
extra variable
$z$ whose power corresponds to the width of the segment under
consideration. We refer to functions including $z$ as ``extended generating
functions''.
We define the extended generating function for irreducible bridges as

$$
X(u,w,z) = P(u,w) + \sum_{m=1}^{\infty} z^{m}T_{m}(u,w)  \eqno (3.2a) $$
and the extended generating function for all bridges as
$$
X^{*}(u,w,z) = P(u,w) + \sum_{m=1}^{\infty} z^{m}T_{m}^{*}(u,w) \eqno (3.2b)
$$

 In these terms, $V(u,w)$, the generating function for bridges ending in
single bonds, generalises to
$$ \tilde{V}(u,w,z) = X(u,w,z) = wz/[1 - wz X(u,w,z)] \eqno (3.3)
$$

 Relation (3.1) generalises to
$$
X^{*}(u,w,z) = X(u,w,z) + X(u,w,z)^{2}\tilde{V}(u,w,z) \eqno (3.4)
$$
whence

$$ X(u,w,z) = X^{*}(u,w,z)/[1 + wzX^{*}(u,w,z)]. \eqno (3.5)
$$

 This relation provides the basis of a suitable truncation.  The
expansion of
(3.5) to order $z^{K}$ requires $T_{m}^{*}$ for $m=1$ to $K$ and will give
$T_{m}$ for
$m=1$ to $K$.  The result, noted above, that $T_{m}$ is of order $w^{3m}$
provides a
useful check on the algebra.

 If we define $R_{m}^{*}(u,w)$ as the generating function for
chains in a strip of
width $m$ such that both ends have the maximal $y$ coordinate then

$$ R_{m}^{*}(u,w) = \frac{1}{2}(P(u,w) - 1) + \sum_{n=1}^{m}
R_{n}(u,w)
\eqno (3.6)$$

 This relation can be easily inverted to give individual
$R_{m}$. These are needed to define $R(u,w)$ in  (2.17)
and also to recover  the $S_m(u,w)$ from $Y(u,w,z)$ (eqn 3.8a)
and   the $Q_m(u,w)$ from $Z(u,w,z)$ (eqn 3.11)

 We define $S_{m}^{*}(u,w)$ as the generating function for chains
in a strip of
width $m$ where one end of the chain has the maximal $y$ coordinate and the
other end
does not have an extremal $y$ coordinate.  The general relation between the
irreducible
and unrestricted generating functions is

$$ S^{*}(u,w) = S(u,w) + [S(u,w) +2 R(u,w)]V(u,w)[P(u,w) +
T(u,w)]  \eqno (3.7)
$$

We define the extended generating function for irreducible components with
one or both ends at the maximal y co-ordinate as

$$ Y(u,w,z) = 2 \sum_{m=1}^{\infty} z^{m}R_{m}(u,w) +
\sum_{m=2}^{\infty}
z^{m}S_{m}(u,w)  \eqno (3.8a)
$$
and the corresponding unrestricted function as

$$ Y^{*}(u,w,z) = 2 \sum_{m=1}^{\infty} z^{m}R_{m}^{*}(u,w) +
\sum_{m=2}^{\infty} z^{m}S_{n}^{*}(u,w), \eqno (3.8b)
$$

Eqn. (3.7) generalises to

$$ Y^{*}(u,w,z) = Y(u,w,z)[1 +
\tilde{V}(u,w,z)X(u,w,z)]  \eqno (3.9)$$
 or
$$ Y(u,w,z) = Y^{*}(u,w,z)/[1 + \tilde{V}(u,w,z)X(u,w,z)] \eqno (3.10)
$$
from which the $S_m$ can be recovered once $R_n$ and $T_n$ are known for $n
\le m$.

 Finally we consider $Q^{*}_{m}$, the generating function for
chains in a strip
$X$ where neither end has an extremal $y$ coordinate.

 We have
$$ Q^{*}(u,w) = Q(u,w) + [P(u,w) +2 R(u,w)]^{2}V(u,w)
\eqno(3.11)$$
Defining
$$ Z(u,w,z) = \sum_{m=2}^{\infty} z^{m}Q_{m}(u,w)
\eqno(3.12a)$$
and
$$ Z^{*}(u,w,z) \sum_{m=2}^{\infty} z^{m}Q^{*}(u,w)
\eqno(3.12b)$$
 gives
$$ Z(u,w,z) = Z^{*}(u,w,z) - Y(u,w,t)^{2} \tilde{V}(u,w,z)
\eqno(3.13)$$

\section{Transfer matrix enumeration techniques}

 The analysis in the previous sections has reduced the problem
of enumerating
general self-avoiding walks of $4K+3$ steps to one of enumerating walks
confined to
strips of width $\leq K$ subject to various constraints.  In order to
enumerate walks
confined to strips we use a transfer matrix technique that generalises the
approach that
we have used in our earlier enumeration of self-avoiding rings.  We draw a
cross-section
line (with a kink) across the width of a strip of width $K$ so as to cut
$K+2$ of the
bonds on which steps of chains can occur.  We note that if we specify the
set of occupied
steps then the self-avoidance constraint acts independently to the left and
right of the
cross-section line.  However not all combinations of self-avoiding
components from the
left and right of the cross-section line combine to give walks.  It is
necessary to
consider the connectivity of the components.  This can be done by
generalising      the
technique that we used in our earlier work.  We assign to each bond
intersected by the
cross-section line an index
$$n_{i} = 0,1,2, \;\quad {\rm or}\; 3  \qquad i = 1 \;\quad {\rm to}\; K +
2.$$

 Here $``0"$ denotes an empty bond, $``1"$ denotes a step
connected to a (uniquely defined)
later step, $``2"$ denotes a step connected to a (uniquely defined) earlier
step and $``3"$
denotes a step not connected to any other steps intersected by the
cross-section line.
\medskip
If we define

$$ A(i,j) = \{k : k \leq j \;\quad {\rm and}\; n_{k} = i\}
\eqno(4.1)$$
 then we require
$$ |A(1,j)| \geq |A(2,j)| \;\quad \hbox{\rm for all}\; j
\eqno(4.2a)$$
$$ |A(1,K+2)| = |A(2,K+2)|
\eqno(4.2b)$$
 as in the enumeration of self-avoiding rings and
$$ |A(3,K+2)| \leq 2.
\eqno(4.2c) $$

 The numbers of sets of $n_{i}$ subject to these constraints for
various $K$ are
given in Table 1.  These numbers give the main limitation on the size of
walks that can be
obtained because it is necessary to store a partial generating function for
the number of
walks corresponding to each allowed set of $n_{i}$.  These numbers,
$s_{k}$, are larger
than the corresponding vector sizes, $r_{k}$, used in the enumeration of
self-avoiding walks, but only by a factor
$\gamma(k)$ which is constrained as

$$ 1 \leq \gamma(k) \leq \frac{1}{2}(k^{2} + 5k + 7)
$$

 Thus the increase in the $s_{k}$ is dominated by a $3^{k/4}$
increase as for
the $r_{k}$.

%

 Self-avoiding chains in strips are developed successively by
advancing the
cross-section line so that one vertex of the lattice passes from the right
to the left of
the line.  Except at the beginning of a column, this corresponds to moving
the kink down
one row.  This move replaces two bonds (and their associated $n_{i}$) by
two new bonds
with new $n_{i}.$  The other $n_{i}$  are unchanged except when the
addition of the new
site changes the connectivity of the components.  Table 2 shows the various
combinations
of new $n_{i}$ that can be produced from various combinations of old
$n_{i}$ pairs.
Various special cases occur.  If a link from a free end (i.e. $n_{i} = 3$)
connects to an
existing loop segment then the other end of the loop must be reset to type
3.  If two
loops meet then one end must be relabelled.  The closing of a single loop
implies that a ring,
disconnected from the walk, has been created and this configuration is
ignored.  The
final step in the construction of a chain is when two type 3 bonds meet.
When this
occurs, all other bonds must be empty for a valid chain generating function
to be added
to the running total.  If this is not the case it implies that other
disconnected
components are present and the configuration is ignored.

 The iteration is initiated from an empty state $(n_{i} \equiv 0)$
with generating
function 1.  As each bond is added, factors of $v$ or $w$ as appropriate are
used to
multiply the old partial generating function and the product is accumulated
into the
running total for the new partial generating function.  Each chain could,
in principle,
be generated for a number of different $x$ and $y$ displacements from a
given reference
origin.  To ensure uniqueness in the $x$ direction, the chain is required
to intersect
the first column considered.  Thus the state with all $n_{i}$ zero is never
continued
after the first column has been built up.  This ensures that each chain is
counted only
once and, together with the requirement that the last operation is to join
two type 3 bonds, also ensures that only one connected component occurs
in each graph that is counted.  Formally,
if $4k + 3$ series terms are required then the transfer matrix operation
must be repeated
until $4k + 3$ columns of each strip have been generated.  This will ensure
that all the
cancellations involved in going from unrestricted to irreducible
contributions will be
correct.  As noted above, the cancellations provide a useful check on the
implementation
of the algebraic formalism.  If however the check is not required then the
results (2.4),
(2.7), (2.10) and (2.14) can be assumed to be true.  For a strip of width
$M$ it is
sufficient to generate only $4k + 3 - 2M$ columns of the strip.

This requires a large amount of memory to store all the intermediate
generating functions. If this amount of memory is not available as
physical memory, but only as virtual (disk based) memory, the
performance can be enhanced greatly by being careful of    the
order in which the partial generating functions are processed.
This can make the entire process close to sequential, and enormously
reduce the amount of disk access required.

As mentioned before, when a partial generating function is processed,
only the two $n_i$ coefficients adjacent to the site being added will
change, unless this changes the connectivity of a loop. However in this
case, it will always be a non-zero going to another non-zero, so
if we make up a ``partial signature'' out of the $n_i$ coefficients
that are not being directly changed, and just record whether they
are zero or not, this partial signature will be invariant under the
transfer matrix. That is, all ``new'' partial generating functions
will have the same ``partial signature''.

We can then process all the partial generating functions with one
particular partial signature, save the results to disk, and process the
next set, until all are processed. We do not need to worry about the
possibility of having to accumulate a new partial generating function
to one stored on disk, as if two partial generating functions have
different partial signatures, they definitely cannot have the same
``full signature'' ($n_i$ values).

The only problem with this is to make sure that we process all the
partial generating functions of a particular partial signature first.
Fortunately, one can get away without having to sort the output
file before using it   as input. If instead of saving
to just one output file, two output files are used, it is possible
to arrange things such that one only has to read from these two files in
order, and the data will be ordered in exactly the right way for
adding the next site. This minimises the amount of disk access.
To obtain this nice ordering, one looks at whether the new value of $n_i$
which will be included in the partial signature for the next phase is
zero or not, and accordingly assigns it to one of the two output files.
This will arrange the partial signatures in binary order, with the
most significant bit in the partial signature being the most recent bit
calculated, and the least significant bit being the oldest bit
calculated.

This method has a side benefit --- in a multi-processor architecture,
one can have each processor working independently on separate
partial signature groups, with very little inter-processor
communication. This means that this algorithm is easy to parallelise,
which will become increasingly important in the future as parallel computers
are becoming more and more popular, whilst many algorithms are
difficult to parallelise.

 The problem of ensuring uniqueness in the $y$-direction requires
inclusion-exclusion
arguments of the type used in our enumeration of self-avoiding rings.  For
a strip of
width $M$ we classify chain generating functions as $G_{m}(\pm,\pm,\pm)$
where $+$ denotes
an allowed location for ends and $-$ denotes a forbidden location.  The
three arguments
refer to the top row, the bottom row and the set of internal rows
respectively. Specifying the end locations permissively, rather than
prescribing
how many end points lie in each set allows us to treat the two ends of the
chain
independently. We have

$$ G_{K}(+,-,-) = {{P-1}\over 2} + \sum_{m=1}^{K} R_{m}
\eqno(4.3a)$$
$$ G_{K}(+,+,-) = T^{*}_{K} + P-1 + 2\sum_{m=1}^{K} R_{m}
\eqno(4.3b)$$
$$ G_{K}(+,-,+) = K{{P-1}\over 2} + \sum_{m=1}^{K} (K+1-m) (R_{m} +
Q^{*}_{m}  + S^{*}_{m}) $$ $$ + \sum_{m=1}^{K} (K-m)(T^{*}_{m} + R_m +
S_m^{*})
\eqno(4.3c)$$

$$ G_{K}(-,-,+) = (K-1){{P-1}\over 2} + \sum_{m=1}^{K} (K+1-m)Q^{*}_{m} +
2 \sum_{m=1}^{K-1} (K-m)(R_{m} + S^{*}_{m}) $$ $$ + \sum_{m=1}^{K-2}
(K-1-m)T^{*}_{m}
\eqno(4.3d)$$

 These relations can be explicitly inverted to give

$$
 R_{m} = G_{m}(+,-,-) - G_{m-1}(+,-,-)
\eqno(4.4a)$$

$$
 Q^{*}_{m} = G_{m}(-,-,+) - G_{m-1}(+,-,+) - \sum_{n=1}^{m-1}
 \left(Q_n^*+R_n+S_n^*\right)
\eqno(4.4b)$$
$$
S^{*}_{m} = G_{m}(+,-,+) - G_{m-1}(+,-,+) - G_{m}(+,-,-) - G_{m-1}(+,-,-)
$$
$$
- {{P-1}\over 2} -Q_m^* - \sum_{n=1}^{m-1}  \left(Q_n^*+2*S_n^*+T_m^*\right)
\eqno(4.4c)$$

$$
 T^{*}_{m} = G_{m}(+,+,-) - 2G_{m}(+,-,-)
\eqno(4.4d)$$

\section {Analysis of series}

This algorithm was implemented for computers in two parts.
The first part was the transfer matrix portion; that is it computed
the $G$ functions. This was the time and memory intensive section. It
was run up to a width of 8 on an IBM RS6000/520 with 256MB or RAM,
then to width 9 on an IBM 3090 with 500MB RAM and 2GB backing store.
It required about 200MB and 600MB of memory respectively, and
required several days in each case. The same machine could have
been used to do width 10 (43 terms) given several   weeks of
processor time. This program was written in C, and
reused memory whenever possible (there was only one bank of memory, used
for both the about-to-be-processed partial generating functions, and
the have-just-been-processed partial generating functions). The method
described previously for using disk storage efficiently was
implemented and tested, but not used as whilst memory was then no longer
a problem, time became a large problem.

The second part performed all the algebra. Whilst algebra on large
(over 15000 coefficients) polynomials in three variables  is slow,
it is still a minor problem compared to the transfer matrix section,
requiring time in only hours and memory in
sub-megabytes, so efficiency was not as vital. It was implemented
in C++.

The results for width 9 (39 terms) are given in table 3.

The method of analysis used is based on first and second order
differential
approximants. It was also used in previous papers [Guttmann (1987),
Guttmann and Wang (1989) and is described
in detail in Guttmann (1989)]. In summary, we construct near-diagonal
inhomogeneous approximants, with the degree of the inhomogeneous polynomial
increasing from 1 to 8 in steps of 1. For first order approximants
(K=1),
 12 approximants are constructed that utilise a given number of series
coefficients, N. Rejecting occasional defective approximants, we form
the mean of the estimates of the critical point and critical exponent
for fixed order of the series, N. The error is assumed to be two
standard deviations. A simple statistical procedure combines the
estimates
for different values of N by weighting them according to the error,
with the estimate with the smallest error having the greatest weight.
As the error tends to decrease with the number of terms used in the
approximant, this procedure effectively weights approximants derived from
a larger number of terms more heavily.

For second order approximants (K=2), we construct 8 distinct
approximants for each value of N. A summary of the results of this
process
is shown in Table 4. The statistical procedure used to combine the
results gives
$$
	x_c = 0.379052 \pm 0.000001 \qquad
	\gamma = 1.3435  \pm 0.0003 \qquad		(K=1)
$$
$$
	 x_c = 0.3790520 \pm 0.0000005 \qquad
        \gamma = 1.3436  \pm 0.00015      \qquad        (K=2)
$$

These results provide abundant support, if support is still needed, for
the value $\gamma = 1.34375$ obtained by Nienhuis (1982, 1984). To refine the
estimate of the critical point, linear regression is used. There is a
strong correlation between estimates of the critical point and critical
exponent. This is quantified by linear regression, and in this way the
biased estimates (biased at $\gamma = 43/32$)
are obtained.

We find
$$
	x_c = 0.3790524 \pm 0.0000005 \qquad  (K=1)
$$
$$
 	x_c = 0.3790525 \pm 0.0000005  \qquad (K=2)
$$

These are in excellent agreement with previous estimates based on the 56
term polygon series (Guttmann and Enting, 1988),
$x_c = 0.37905228 \pm 0.00000014$

For the honeycomb lattice, the ``connective constant'' $= 1/x_c$ is known
exactly (Nienhuis 1982, 1984),
and is $\sqrt{2 + \sqrt{2}}$, which satisfies a simple
quadratic equation in $x_c^{2}$. A feature of Maple (Version 5) is a clever
algorithm for seeking polynomials with integer coefficients that have
a given root. Attempting to find a quartic polynomial that gave as a
root the biased value of $x_c$ quoted above, we found the best solution
was also a polynomial quadratic in $x_c^{2}$. It was
	$$581x^4 + 7x^2 - 13  = 0.$$

The root is $x_c = 0.37905227... $. While we consider it would
be fortuitous if this were the true value of the critical point,
it nevertheless provides a useful mnemonic.

Another analysis we were able to carry out with this long series was a study
of amplitudes of the leading term and the correction terms. As previously
discussed for self-avoiding polygons (Guttmann and Enting 1988),
 we have found no evidence for
any non-analytic correction-to-scaling term other than
that suggested by Nienhuis, with a ``correction''
exponent of $\Delta = 1.5$. In the case of the polygon generating function
 this ``folds into''
the additive analytic term. However, for the SAW series, it gives rise to
a non-analytic correction term. Furthermore, there is another singularity on
the
negative real axis, at $x = -x_c$, as shown by Guttmann and Whittington
(1978).

Thus we expect the generating function for walks to behave like
$$
	C(x) = \Sigma c_nx^n \sim A(x)(1-\mu x)^{-43/32}[1 + B(x)(1-\mu x)^{3/2} ...
]
                + D(x)(1+\mu x)^{-1/2}. \eqno{(5.1)}
$$

 The exponent for the singularity on the negative real axis reflects the
fact that this term is expected to behave as the energy, and hence to have
exponent
$\alpha - 1$, where $\alpha = {{1}\over{2}}$. From the above, it follows that
the
 asymptotic form of the
coefficients, $c_n$, behaves like:
$$
	c_n \sim \mu^n[a_1n^{11/32} + a_2n^{-21/32} + b_1n^{-37/32}
        + (-1)^nd_1n^{-3/2}+(-1)^nd_2n^{-5/2}] \eqno{(5.2)}
$$

The five amplitudes, $a_1, a_2, b_1, d_1$ and $d_2$ come from the leading
singularity
(giving rise to $a_1$ and $a_2$), the correction-to-scaling term (giving rise
to $b_1$)
and the term on the negative real axis (giving rise to $d_1$ and $d_2$). A
small program
written in Mathematica was used to fit successive quintuples of coefficients,
$c_{n-4}, c_{n-3}, c_{n-2}, c_{n-1}$ and $c_n$ for $n$ = 6,7,8,...,39. The
results
are given in table 5.

With the possible exception of the sequence $\{d_2\}$, the sequences for the
various
amplitudes appear to be converging. Various other values for the exponents
were
also tried, including a square-root correction-to-scaling term. In all cases
the convergence was dramatically worsened by such changes. Indeed, with a
square-root correction-to-scaling exponent, a number of sequences appeared to
diverge rather than converge. However, we have assumed above that the
sub-leading
 term of the singularity on the negative real axis is analytic. If
we allow this singularity to be a square root singularity, so that the last
term in eqn. (5.2) above becomes $\mu^n(-1)^nd_2n^{-2}$ then the results are
even better converged, as shown in Table 6.

 From these tables we estimate $a_1 \approx 1.1771, a_2 \approx 0.554, b_1
\approx
-0.19, d_1 \approx -0.19$, where errors are expected to be confined to the
last
quoted digit in each case. Repeating the above calculations with a critical
point
shifted by twice the confidence limit quoted does not change these amplitude
estimates.

This then completes our numerical study of the generating function for
self avoiding walks.

\ack

One of us (A.R.C.) would like to thank the A.O. Capell, Wyselaskie
and Daniel Curdie scholarships.
We would like to acknowledge the support of ACCI and The University
of Melbourne for the provision of the computers on which these
calculations were performed. Mr Glenn Wightwick of IBM (Australia)
provided
valuable help in running large jobs on the IBM 3090, and we wish to
express our thanks for this assistance. This work was supported by
 the Australian
Research Council.

 \references
\refjl{Enting I G , 1980}{  J. Phys. A: Math. Gen}{13}{3713.}

\refjl{Enting I G  and Guttmann A J  1985}{\JPA}{18}{1007.}

\refjl{\dash  1989}{\JPA}{22}{1371}

\refjl{\dash 1990}{ J Stat. Phys.}{58}{475}

\refjl{ \dash 1992}{\JPA}{25}{2791}

\refjl{Guttmann A J, 1987}{\JPA}{20}{1839}

\refbk{\dash 1989 In:}{Phase Transitions and Critical Phenomena}{ 13.  eds.
C. Domb and J. Lebowitz (Academic Press, London)}

\refjl{Guttmann A J  and Enting I G , 1988}{\JPA}{21}{L165}

\refjl{Guttmann A J and Wang J S, 1991}{\JPA}{24}{3107}

\refjl{Guttmann A J and Whittington S G 1978}{\JPA}{11}{721}

\refjl{MacDonald D, Hunter D L, Kelly K and Jan N, 1992}{\JPA}
{25}{1429}

\refjl{Masand B, Wilensky U, Maffar J P and Redner S,
 1992}{\JPA}{25}{L365}

\refjl{Nienhuis B, 1982}{Phys. Rev. Lett.}{49}{1062}

\refjl{Nienhuis B, 1984}{J. Stat. Phys.}{34}{731}

\refjl{Sykes M F and Guttmann A J, Roberts P D and Watts M G 1972}{\JPA}
{5}{653}

\tables

\tabcaption{
The sizes of vectors required by the transfer matrix formalism.  For ring
enumeration
$r_{k}$ components are required.  For walk enumeration $s_{k}$ components
are required.}

\boldrule{5.5truein}
\halign{#\hfil&\hfil\qquad#&\hfil\quad#&\hfil\quad#\cr
Strip width, k & \#bonds & $r_{k}$ & $s_{k}$ \cr
\noalign{\medrule{5.5truein}  }
-     & 1     & 1     & 2 \cr
-     & 2     & 2     & 5 \cr
1     & 3     & 4     & 13 \cr
2     & 4     & 9     & 37 \cr
3     & 5     & 21     & 106 \cr
4     & 6     & 51     & 312 \cr
5     & 7     & 127     & 925 \cr
6     & 8     & 323     & 2767 \cr
7     & 9     & 835     & 8314 \cr
8     & 10     & 2188     & 25073 \cr
9     & 11     & 5798     & 75791 \cr
10     & 12     & 41835     & 229495 \cr }
\boldrule{5.5truein}

\vfill

\eject

\tabcaption{Allowed transformations of bond indices. For the operations,
``Build''
means incorporate the contribution into the new vector, as defined by the new
indices. ``R$(a \rightarrow b)$'' means apply the change $a \rightarrow b$ to
the
other end of the chain in order to specify the index in the new vector.
``Ignore''
means perform no operation, as a disconnected ring has been generated. ``Acc''
means accumulate the vector component into the chain generating function  if
all the other $n_i$ are zero, otherwise the operation preceding the ``OR'' is
applied}

\boldrule{6.5truein}
\halign{#\hfil&\hfil\qquad#&\hfil\quad#\cr
Old indices  & New indices & \quad Operation  \cr
$(n_{j},n_{j+1})$ & $(n_{j},n_{j+1})$ & \cr
\noalign{\medrule{6.5truein}}
(0,0) & (0,0), (0,3), (3,0) {\rm or} (1,2) &
 \quad {\rm Build}\cr
(0,1)  \quad {\rm or}  (1,0) & (0,1), (1,0)  \quad {\rm or}  (0,0) &
\quad  R(2 $\rightarrow$ 3)
\cr
 (0,2)  \quad {\rm or}  (2,0) & (0,2), (2,0)  {\rm or}  (0,0) &
\quad R(1 $\rightarrow$ 3)
\cr
  (0,3)  \quad {\rm or} (3,0) & (0,3),
(3,0) &
\quad {\rm Build OR Acc}
\cr
 (1,1) & (0,0) & \quad R(2 $\rightarrow$ 1)
\cr
(2,2) & (0,0) &
\quad R(1 $\rightarrow$ 2)
\cr
(1,2) & - & \quad {\rm Ignore}
\cr
(2,1) & (0,0) &
\cr
(3,3) & - & \quad {\rm  Ignore OR Acc}
 \cr
(3,1), (1,3) & (0,0) & \quad R(2 $\rightarrow$ 3)
 \cr
(3,2), (2,3) & (0,0) &
 \quad R(1 $\rightarrow$ 3)
  \cr}
\boldrule{6.5truein}

\vfill

\eject
\baselineskip=10pt
\tabcaption{
Numbers of self avoiding walks.
}

\boldrule{5.5truein}
\halign{#\hfil&\hfil\quad#\cr
$n$ & $ c_n$ \cr
\noalign{\medrule{5.5truein}  }
0 & 1 \cr
1 & 4 \cr
2 & 12 \cr
3 & 36 \cr
4 & 100 \cr
5 & 284 \cr
6 & 780 \cr
7 & 2172 \cr
8 & 5916 \cr
9 & 16268 \cr
10 & 44100 \cr
11 & 120292 \cr
12 & 324932 \cr
13 & 881500 \cr
14 & 2374444 \cr
15 & 6416596 \cr
16 & 17245332 \cr
17 & 46466676 \cr
18 & 124658732 \cr
19 & 335116620 \cr
20 & 897697164 \cr
21 & 2408806028 \cr
22 & 6444560484 \cr
23 & 17266613812 \cr
24 & 46146397316 \cr
25 & 123481354908 \cr
26 & 329712786220 \cr
27 & 881317491628 \cr
28 & 2351378582244 \cr
29 & 6279396229332 \cr
30 & 16741957935348 \cr
31 & 44673816630956 \cr
32 & 119034997913020 \cr
33 & 317406598267076 \cr
34 & 845279074648708 \cr
35 & 2252534077759844 \cr
36 & 5995740499124412 \cr
37 & 15968852281708724 \cr
38 & 42486750758210044 \cr
39 & 113101676587853932 \cr }
\boldrule{5.5truein}

\vfill

\eject

\tabcaption{
Estimates of the critical point ($x_c$) and critical exponent ($\gamma$)
from first order (K=1) and second order (K=2) differential approximants.
$L$ is the number of approximants used. If $L$ is too small (marked with
an `x'), the estimates are not used in the subsequent statistical
analysis.
}

\boldrule{5.5truein}
\halign{#\hfil&\hfil\quad#&\hfil\quad#&\hfil\quad#&\hfil\quad#&\hfil\quad#&\hfil\quad#\cr
$K$ & $n$ & $x_c$ & error & $\gamma$ & error & $L$ \cr
\noalign{\medrule{5.5truein}  }
1& 19 & .3790473 & --- & --1.3430184 & --- & 1x\cr
1& 20 & .3790495 & .0000004 & --1.3432502 & .0001073 & 2x\cr
1& 21 & .3790526 & .0000105 & --1.3435231 & .0016130 & 4 \cr
1& 22 & .3790469 & .0000133 & --1.3427806 & .0021290 & 4 \cr
1& 23 & .3790468 & .0000071 & --1.3427666 & .0011925 & 5 \cr
1& 24 & .3790520 & .0000038 & --1.3436373 & .0006432 & 7 \cr
1& 25 & .3790525 & .0000052 & --1.3437004 & .0010102 & 9 \cr
1& 26 & .3790508 & .0000050 & --1.3433692 & .0010094 & 11 \cr
1& 27 & .3790530 & .0000074 & --1.3437617 & .0013817 & 12 \cr
1& 28 & .3790519 & .0000016 & --1.3435724 & .0003713 & 12 \cr
1& 29 & .3790519 & .0000010 & --1.3435632 & .0002717 & 9 \cr
1& 30 & .3790517 & .0000016 & --1.3435091 & .0004340 & 9 \cr
1& 31 & .3790518 & .0000011 & --1.3435230 & .0003276 & 8 \cr
1& 32 & .3790514 & .0000017 & --1.3434149 & .0005073 & 4 \cr
1& 33 & .3790521 & .0000016 & --1.3436098 & .0004159 & 9 \cr
1& 34 & .3790525 & .0000028 & --1.3437270 & .0007447 & 11 \cr
1& 35 & .3790518 & .0000003 & --1.3435417 & .0000956 & 10 \cr
1& 36 & .3790519 & .0000003 & --1.3435722 & .0001233 & 10 \cr
1& 37 & .3790517 & .0000011 & --1.3434730 & .0004568 & 8 \cr
1& 38 & .3790518 & .0000009 & --1.3435048 & .0003501 & 9 \cr
1& 39 & .3790521 & .0000001 & --1.3436392 & .0000515 & 2x\cr
\noalign{\medrule{5.5truein}  }
2& 28 & .3790525 & -- & --1.3437307 & -- & 1x\cr
2& 29 & .3790520 & -- & --1.3435885 & -- & 1x\cr
2& 30 & .3790518 & .0000006 & --1.3435431 & .0001628 & 2x\cr
2& 31 & .3790518 & .0000005 & --1.3435300 & .0001315 & 3x\cr
2& 32 & .3790513 & .0000016 & --1.3432041 & .0010873 & 3x\cr
2& 33 & .3790515 & .0000007 & --1.3433885 & .0004434 & 4 \cr
2& 34 & .3790519 & .0000003 & --1.3435503 & .0000824 & 5 \cr
2& 35 & .3790521 & .0000004 & --1.3436142 & .0001314 & 5 \cr
2& 36 & .3790519 & .0000006 & --1.3435607 & .0001972 & 6 \cr
2& 37 & .3790520 & .0000001 & --1.3435822 & .0000276 & 7 \cr
2& 38 & .3790520 & .0000001 & --1.3435845 & .0000264 & 6 \cr
2& 39 & .3790521 & .0000002 & --1.3436174 & .0000616 & 4 \cr
} \boldrule{5.5truein}

\vfill

\eject

\tabcaption{
Sequences of amplitude estimates. Refer equation (5.2)
}

\boldrule{5.5truein}
\halign{#\hfil&\hfil\qquad#&\hfil\quad#&\hfil\quad#&\hfil\quad#&\hfil\quad#\cr
$n$ & $ d_2$ &  $d_1$ & $b_1$ & $a_2$ & $a_1$ \cr
\noalign{\medrule{5.5truein}  }
29 &    0.0639& --0.1878& --0.1999& 0.5584& 1.17700\cr
30 &    0.0666& --0.1879& --0.2022& 0.5590& 1.17699\cr
31 &    0.0715& --0.1881& --0.1980& 0.5579& 1.17700\cr
32 &    0.0738& --0.1882& --0.1999& 0.5584& 1.17700\cr
33 &    0.0781& --0.1883& --0.1963& 0.5574& 1.17701\cr
34 &    0.0800& --0.1884& --0.1979& 0.5578& 1.17700\cr
35 &    0.0838& --0.1885& --0.1947& 0.5570& 1.17701\cr
36 &    0.0855& --0.1885& --0.1960& 0.5573& 1.17701\cr
37 &    0.0890& --0.1886& --0.1932& 0.5566& 1.17701\cr
38 &    0.0904& --0.1887& --0.1943& 0.5569& 1.17701\cr
39 &    0.0936& --0.1888& --0.1919& 0.5563& 1.17702\cr }
\boldrule{5.5truein}

\vfill

\tabcaption{
Sequences of amplitude estimates, with the exponent associated with
$d_2$ changed from $-{5\over 2}$ to $-2$. Refer equation (5.2)
}

\boldrule{5.5truein}
\halign{#\hfil&\hfil\qquad#&\hfil\quad#&\hfil\quad#&\hfil\quad#&\hfil\quad#\cr
$n$ & $ d_2$ &  $d_1$ & $b_1$ & $a_2$ & $a_1$ \cr
\noalign{\medrule{5.5truein}  }
29& 0.0246& --0.1902& --0.2004& 0.5585& 1.17699 \cr
30& 0.0252& --0.1903& --0.2017& 0.5589& 1.17699 \cr
31& 0.0266& --0.1906& --0.1985& 0.5580& 1.17700 \cr
32& 0.0270& --0.1906& --0.1994& 0.5583& 1.17700 \cr
33& 0.0281& --0.1908& --0.1969& 0.5576& 1.17700 \cr
34& 0.0283& --0.1909& --0.1974& 0.5577& 1.17700 \cr
35& 0.0292& --0.1910& --0.1952& 0.5571& 1.17701 \cr
36& 0.0293& --0.1910& --0.1955& 0.5572& 1.17701 \cr
37& 0.0301& --0.1912& --0.1937& 0.5568& 1.17701 \cr
38& 0.0301& --0.1912& --0.1939& 0.5568& 1.17701 \cr
39& 0.0308& --0.1912& --0.1923& 0.5564& 1.17702 \cr }
\boldrule{5.5truein}

\eject

\figures


\figcaption{Schematic representation
of projections of self-avoiding walks onto the $y$ axis.}

\figcaption{The 5 types of irreducible
 component from which self-avoiding walks
are constructed.}

\figcaption{The cross-section line
 defining the set of lattice bonds which
specify the partial generating functions.}

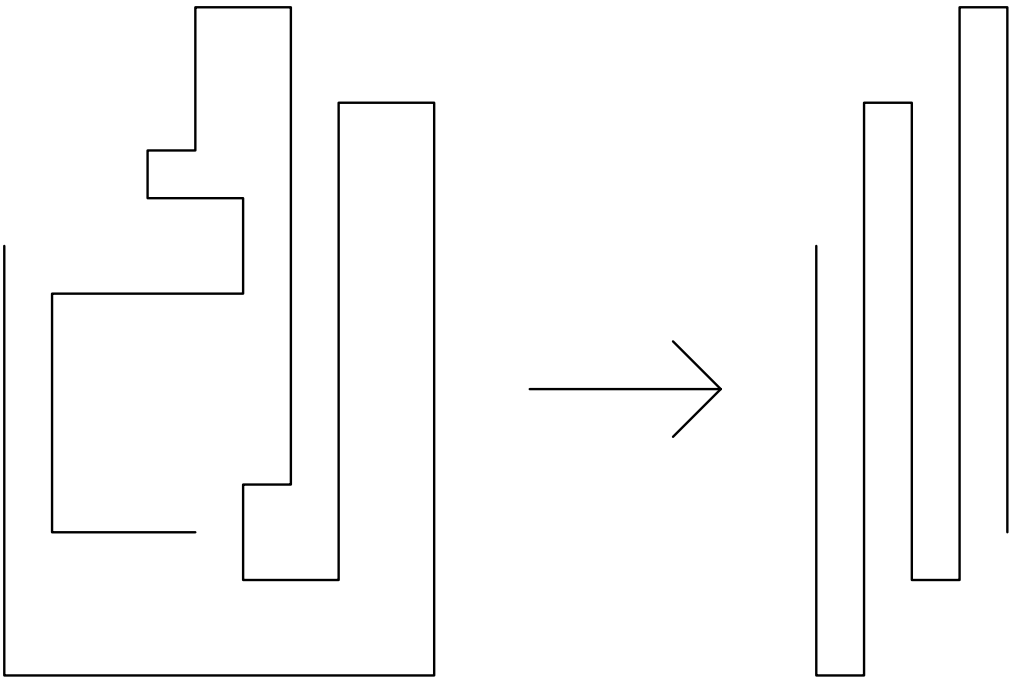
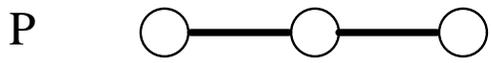
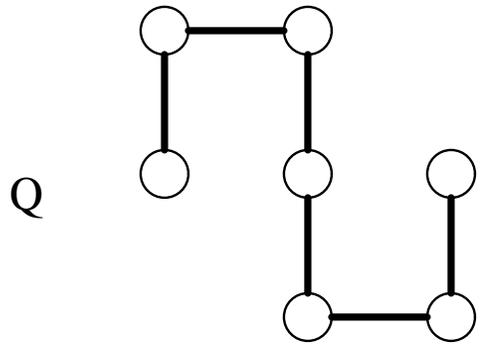
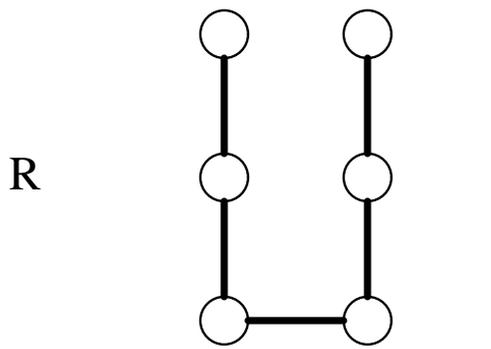
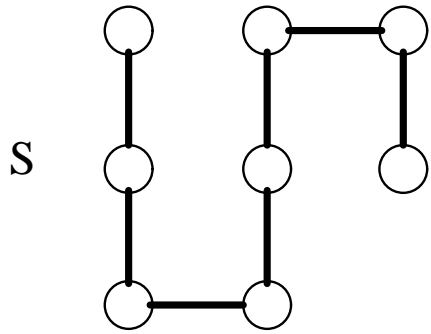
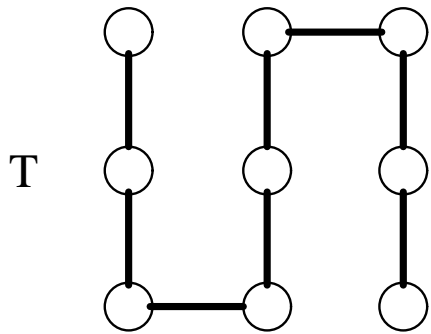
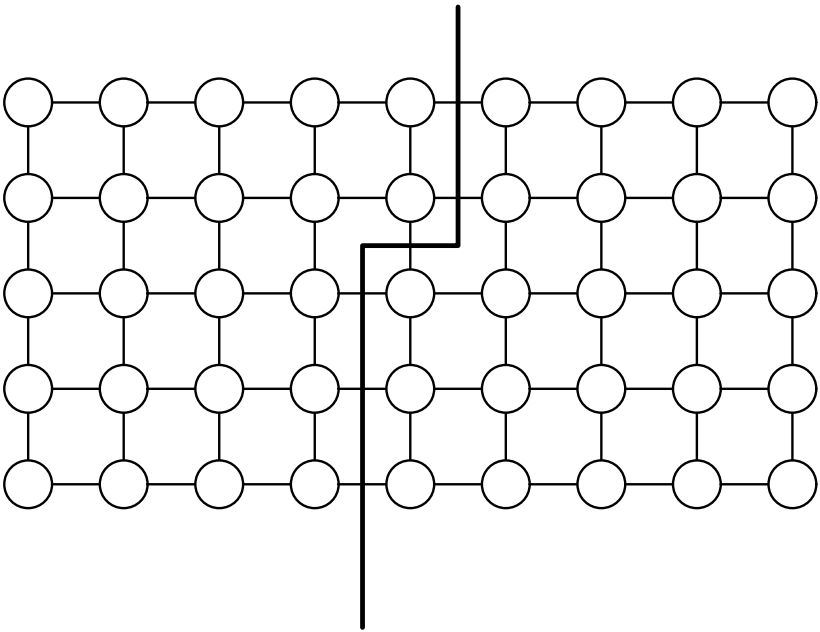
\bye